\documentclass[preprint,12pt]{elsarticle}
\usepackage{amssymb}
\usepackage{amsmath}

\journal{Chaos Solitons and Fractals}
\input{tcilatex}
\begin{document}

\begin{frontmatter}



\title{Trajectory statistics and turbulence evolution}


\author{M. Vlad and F. Spineanu}

\address{National Institute of Laser, Plasma and Radiation Physics \\
Atomistilor 409, 077125 Magurele, Bucharest, Romania}

\begin{abstract}
The aim of this paper is to understand the tendency to organization of the turbulence in two-dimensional ideal fluids. We show that nonlinear processes as inverse cascade of the energy and vorticity concentration are essentially determined by trajectory trapping or eddying. The statistics of the trajectories of the vorticity elements is studied using a semianalytic method. The separation of the positive and negative vorticity is due to the attraction produced by a large scale vortex on the small scale vortices of the same sign. More precisely, a large scale velocity is shown to determine average transverse drifts, which have opposite orientations for positive and negative vorticity. They appear in the presence of trapping and lead to energy flow to large scales due to the increase of the circulation of the large vortex.
Recent results on drift turbulence evolution in magnetically confined plasmas are discussed in order to underline the idea that there is a link between the inverse cascade and trajectory trapping. The physical mechanisms are different in fluids and plasmas due to the different types of nonlinearities of the two systems, but trajectory trapping has the main role in both cases.  

\end{abstract}

\begin{keyword}
test particle statistics, turbulence, self-organization, vorticity, Euler fluid


\end{keyword}

\end{frontmatter}



\section{Introduction}

The two-dimensional fluid turbulence represented during more than 70 years
an active field of research (see the review papers \cite{Kraichnan1980}, 
\cite{Provenzale1999}, \cite{Kukharkin1995} and the references therein). It
has many applications in different areas as fluid dynamics, meteorology,
oceanography, fusion plasmas, superfluids, superconductors and astrophysics,
in spite of the fact that it provides only idealized models for physical
systems that are always three-dimensional.

The two-dimensional turbulence has a self-organizing character, which is
related to the invariance of both energy and enstrophy in ideal (inviscid)
fluids. Numerical studies of the decaying turbulence clearly show this
property and the associated scaling behaviour \cite{McWilliams84}, \cite%
{Weiss1992}, \cite{Maassen2002}. The enstrophy has a direct cascade (to
small scales), but with a complex evolution characterized by the presence of
inverse cascade in isolated regions \cite{Ohkitani}, \cite{Babiano}. The
energy has an inverse cascade (to large scales) that leads to the emergence
of large quasi coherent vortices. The process of self-organization can
continue until the coherent vortices reach the size of the system \cite%
{Matthaeus1991}, \cite{Montgomery1992}. This behaviour was explained in the
representation of point-like vortices by a negative temperature (\cite%
{Eduards1974}, \cite{Kraichnan1980}, \cite{Robert1992}) or by the property
of self-duality of the associated field theoretical model (\cite%
{Spineanu2003}). The latter approach was extended to a model of planetary
atmosphere and magnetized plasmas \cite{Spineanu2005}.

This paper deals with the turbulent states and studies the self-organization
during its initial stage, before the emergence of large coherent vortices of
the system size.

We show that trajectory trapping or eddying in the structure of the
turbulence is the main physical reason for the strong nonlinear effects that
are observed in two-dimensional ideal fluids. This conclusion is drawn from
a study of the statistics of test particles (tracers) in turbulent Euler
fluids.

This study is based on a series of recent results on the statistical
properties of test particle trajectories in incompressible two-dimensional
velocity fields. Numerical simulations have shown that trajectories are
complex, as they have both random and quasi-coherent aspects. A typical
trajectory, shown in Figure 1, is a random sequence of long jumps and
trapping events that consists of winding on almost closed paths. Analytical
methods that describe the statistics of these trajectories were developed 
\cite{Vlad1998}, \cite{Vlad2004} and used for understanding various aspects
of turbulent transport. It was shown that they provide a very good
description of the nonlinear effects produced by trajectory trapping or
eddying and reasonably accurate quantitative results for the diffusion
coefficients and for other statistical averages.

The conclusion of these studies is that trajectory trapping or eddying leads
to nonstandard statistics: memory effects (represented by long time
Lagrangian correlation), strongly modified transport coefficients and
non-Gaussian distributions of displacements. It was also shown that trapping
determines a large degree of coherence in the sense that bundles of
trajectories that start from neighboring points remain close for very long
time compared to the eddying time. Trapped trajectories form quasi-coherent
structures similar to fluid vortices. Extensive theoretical \cite{Vlad2003}-%
\cite{NeurSpatscheck2006} and numerical studies \cite{Zimbardo2000}-\cite%
{Beresnyak2011} have contributed significantly in the last decades to the
understanding of the turbulent transport in laboratory or space plasmas, in
fluids or in stochastic magnetic fields.

The connection between the test particle trapping and the turbulence
evolution was discussed in \cite{Vlad2013} based on a study of test modes on
turbulent plasmas. Analytical results that are in agreement with numerical
simulations were obtained, and they allowed to deduce a new physical
perspective on the nonlinear process of generation of large scale
correlations (inverse cascade) and of zonal flow modes. Essentially, they
are effects of ion trajectory trapping or eddying.

We show here that trapping has an essential role in two-dimensional fluid
turbulence. A nonlinear effect produced by trapping of the vorticity
elements explains the separation of positive and negative vorticity and the
inverse cascade of the energy.

The paper is organized as follows. The problem of test particle or tracer
transport is defined in Section 2.1 and Section 2.2 contains a short
presentation of the analytical statistical approach, the decorrelation
trajectory method (DTM). The effects of trapping or eddying on tracer
transport and on the statistical characteristics of the trajectories are
discussed in Section 3. This section contains a review of the previous work
and new results on the modifications of the trajectory structures determined
by an average velocity. The effects of trapping on the decaying
two-dimensional turbulence in ideal fluids are analyzed in Section 4.1. The
physical process that determines the separation of the positive and negative
vorticity and leads to the inverse cascade of the energy is identified. The
average speed of vorticity separation is estimated. Section 4.2 is a short
discussion on recent results on plasma turbulence evolution, which show that
trajectory trapping has an essential role in the inverse cascade, although
the physical mechanism is completely different. The conclusions are
summarized in Section 5.

\section{Test particle transport and the statistical method}

\subsection{The problem}

The problem of test particle or tracer advection in two-dimensional
incompressible velocity fields is described by the stochastic equation: 
\begin{equation}
\frac{d\mathbf{x}(t)}{dt}=\mathbf{v}\left[ \mathbf{x}(t),t\right] +V_{d}%
\mathbf{e}_{y},  \label{1}
\end{equation}%
where $\mathbf{x}(t)$ represents the trajectory in the plane $(\mathbf{e}%
_{x},~\mathbf{e}_{y})$ and $V_{d}\mathbf{e}_{y}$ is an average velocity that
is taken in the $\mathbf{e}_{y}$ direction. The stochastic velocity $\mathbf{%
v}(\mathbf{x},t)$ is incompressible $[\mathbf{\nabla \cdot v}(\mathbf{x},t)=$
$\mathbf{0]}$ and is represented by a scalar field, the stochastic potential
or stream function%
\begin{equation}
\mathbf{v}(\mathbf{x},t)=\mathbf{e}_{z}\times \mathbf{\nabla }\phi
=(-\partial _{y}\phi ,\ \partial _{x}\phi ).  \label{v}
\end{equation}%
The potential $\phi (\mathbf{x},t)$ is considered to be a stationary and
homogeneous Gaussian stochastic field, with zero average and given two-point
Eulerian correlation function (EC) 
\begin{equation}
E(\mathbf{x},t)\equiv \left\langle \phi (\mathbf{x}^{\prime },t^{\prime
})\,v_{j}(\mathbf{x}^{\prime }+\mathbf{x},t^{\prime }+t)\right\rangle
\label{pec}
\end{equation}%
where $\left\langle ...\right\rangle $ denotes the statistical average over
the realizations of $\phi (\mathbf{x},t)$ or the integral over $\mathbf{x}%
^{\prime }$ and $t^{\prime }.$\ The main parameters of the EC\ are: the
amplitude of the potential fluctuations $\beta ^{2}=E(\mathbf{0},0),$ the
correlation length $\lambda _{c}$\ and the correlation time $\tau _{c},$\
which are the characteristic length and time of the decay of the function $E(%
\mathbf{x},t).$\ The EC's of the velocity components are obtained as space
derivatives of $E(\mathbf{x},t)$ and the amplitude of the stochastic
velocity is $V=\beta /\lambda _{c}.$

Starting from the above statistical description of the stochastic potential
and from an explicit EC one has to determine the statistical properties of
the trajectories. This problem is nonlinear due to the space dependence of
the potential, which leads to $\mathbf{x}$ dependence of the EC (\ref{pec}).

The trajectories are solutions of a Hamiltonian system and thus, for time
independent potential $\phi (\mathbf{x}),$ their paths are the contour lines
of the total potential $\phi (\mathbf{x})+xV_{d}$ (the stream lines). The
velocity is a continuous function of $\mathbf{x}$ and $t$ in each
realization and it determines an unique trajectory as the solution Eq. (\ref%
{1}) with the initial condition $\mathbf{x}(0)=\mathbf{0.}$

For $V_{d}=0$ the stream lines are closed curves (with the exception of one
contour line, $\phi (\mathbf{x})=0,$\ which extends to infinite), and the
trajectories are periodic functions of time. They correspond to permanent
trapping. This invariance property is approximately maintained for
potentials that are weakly perturbed by a slow time variation. The
trajectories approximately follow the contour lines of $\phi (\mathbf{x},t)$
on almost closed paths for time intervals that are larger than the time of
flight $\tau _{fl}=\lambda _{c}/V.$ This leads to typical trajectories
similar to the example shown in Figure 1. They have a complex structure that
consists of a random sequence of trapping events and long jumps. Trapping
appears at different scales when the particles are close to the maxima or
minima of the potential. The large displacements are produced when the
trajectories are at small absolute values of the potential. The importance
of trapping is measured by the Kubo number defined by

\begin{figure}[h]
\centerline{\includegraphics[height=6cm]{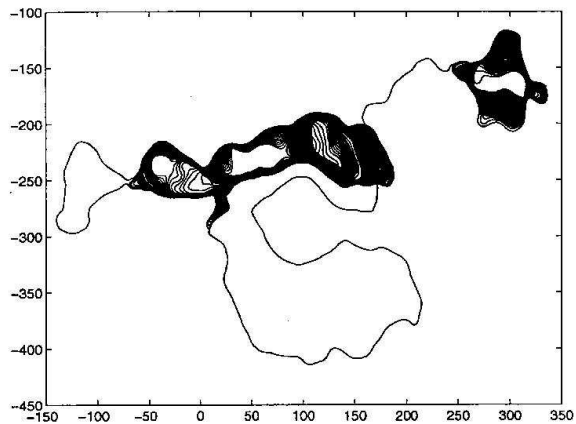}}
\caption{A typical trajectory obtained from Eq.(1) for$K=10,$ $V_d=0$ and
the stream function used in the simulations presented in [30].}
\end{figure}

\begin{equation}
K\equiv \frac{\tau _{c}}{\tau _{fl}}=\frac{V\tau _{c}}{\lambda _{c}}\,.
\label{K}
\end{equation}%
Trajectory trapping appears for $K>1$ and becomes stronger as $K$ increases
up to the limit of static or frozen fields ($K,$ $\tau _{c}=\infty )$ when
the trapping is permanent.

An average velocity $V_{d}$ changes the topology of the total potential $%
\phi (\mathbf{x})+xV_{d}$ by generating bunches of opened contour lines
along its direction. When $r\equiv V_{d}/V<1,$ islands of closed contour
lines persist between the bunches of opened lines, which enables trajectory
trapping. At large average velocities $r\gg 1$ all the contour lines are
opened and test particle cannot be trapped. The average velocity defines a
characteristic time for traversing the correlation length with the average
velocity, $\tau _{\ast }=\lambda _{c}/V_{d},$\ and a dimensionless
parameter\ similar to the Kubo number%
\begin{equation}
K_{\ast }\equiv \frac{\tau _{\ast }}{\tau _{fl}}=\frac{V}{V_{d}}=\frac{1}{r}.
\label{Kstar}
\end{equation}

The condition that trapped trajectories are obtained from Eq. (\ref{1}) is $%
K>1$ and $K_{\ast }>1$ (or $r<1).$

\bigskip

Test particle transport is essentially determined by the Lagrangian velocity
correlation (LVC) defined by

\begin{equation}
L_{ij}(t)\equiv \left\langle v_{i}\left[ \mathbf{x}(0),0\right] \ v_{j}\left[
\mathbf{x}(t),t\right] \right\rangle  \label{CL}
\end{equation}%
for a stationary process. As shown by Taylor \cite{Taylor}, the mean square
displacement $\left\langle x_{i}^{2}(t)\right\rangle $ and its derivative,
the running diffusion coefficient $D_{i}(t)$, are integrals of the LVC%
\begin{equation}
\left\langle x_{i}^{2}(t)\right\rangle =2\int_{0}^{t}d\tau \;L_{ii}(\tau
)\;(t-\tau ),  \label{MSD}
\end{equation}%
\begin{equation}
D_{i}(t)=\int_{0}^{t}d\tau \;L_{ii}(\tau ).  \label{D}
\end{equation}%
The asymptotic values of $D_{i}(t)$\ 
\begin{equation}
\chi _{i}=\underset{t\rightarrow \infty }{\lim }D_{i}(t)=\int_{0}^{\infty
}L_{ii}(\tau )d\tau  \label{das}
\end{equation}%
are the diffusion coefficients. The time dependent diffusion coefficients
provide\ the "microscopic" characteristics of the transport process. The
diffusion at the transport space-time scales, which are much larger than $%
\lambda _{i}$ and $\tau _{c},$ is described by the asymptotic values $\chi
_{i}.$

The scaling of the diffusion coefficient in the Kubo number can be obtained
by a simple estimation based on the general shape of the LVC (\ref{CL}). It
is usually a function that decays to zero from the value $V^{2}$ at $t=0$ in
a characteristic time $\tau _{d}$ (the decorrelation time). For small Kubo
numbers $K\ll 1$ $(\tau _{c}\ll \tau _{fl}),$ the time variation of the
velocity field is fast and the particles cannot \textquotedblright
see\textquotedblright\ the space structure of the velocity field. In this
quasilinear regime (or the weak turbulence case) $\tau _{d}=\tau _{c}$ and
the diffusion coefficient (\ref{das}) is $\chi _{ql}\approx V^{2}\tau _{c}=($
$\lambda _{c}^{2}/\tau _{c})K^{2}.$ In the nonlinear regime $K\gg 1$ $(\tau
_{fl}\ll \tau _{c}),$ the decay time of the LVC is $\tau _{d}=\tau _{fl}$
and the diffusion scales as the Bohm diffusion coefficient $\chi _{B}\approx
V^{2}\tau _{fl}=($ $\lambda _{c}^{2}/\tau _{c})K,$ which does not dependent
on $\tau _{c}.$ The shape of the EC determines only a numerical factor in
the diffusion coefficient.

The Bohm scaling does not apply in the case of diffusion in 2-dimensional
divergence-free velocity fields described by Eq. (\ref{1}). Numerical
simulations have shown that the diffusion coefficient does not saturate as $%
\tau _{c}$ increases, but it decreases with the increase of the correlation
time at $K>1\ (\tau _{c}>\tau _{fl}).$ Trajectory trapping or eddying was
found to produce this effect \cite{kraichnan}. The scaling in the nonlinear
regime is%
\begin{equation}
\chi _{tr}\approx (\lambda _{c}^{2}/\tau _{c})K^{\alpha }=\beta K^{\alpha
-1},  \label{dtr}
\end{equation}%
where $\alpha <1.$ In particular, in the limit of static or frozen potential 
$K\rightarrow \infty ,$ the process is subdiffusive with $\chi
_{tr}\rightarrow 0.$\ \ \ 

\subsection{The decorrelation trajectory method (DTM)}

Several methods (as Corrsin approximation \cite{Corrsin1959}, \cite{McComb},
direct interaction approximation \cite{Roberts}, the renormalization group
technique \cite{bouchaud}) were developed for determining the LVC
corresponding to given EC. All these methods lead to finite diffusion
coefficient of Bohm type in the limit $K\rightarrow \infty ,$ which shows
that they are not adequate for the two-dimensional incompressible velocity
fields. This process was studied especially by means of direct numerical
simulations (\cite{jacques} and the reference there in) or by developing
simplified models \cite{majda}. There is a theoretical estimation \cite%
{isichenko} based on the analogy with the fractal structure of the
landscapes. Using the theory of percolation, it finds the scaling of the
diffusion coefficient of the type (\ref{dtr}) with $\alpha =0.7.$ It is the
first analytical estimation that has obtained subdiffusion in the frozen
turbulence. The problem is that this method can provide only the scaling of $%
\chi ,$ but not the statistics of the test particle trajectories.

Important analytical results in the study of this special case were obtained
in the last decades by developing new statistical methods (the decorrelation
trajectory method DTM \cite{Vlad1998} and the nested subensemble approach
NSA \cite{Vlad2004}). They determine the LVC, the time dependent diffusion
coefficients $D_{i}(t),$ the probability of displacements until
decorrelation and other statistical averages. It was shown that the presence
of trapping determines memory effects in $L(t)$ and a rich class of
anomalous diffusion regimes in the presence of a decorrelation mechanism 
\cite{Vlad2004}. The trapping has also collective effects. It determines
coherence in the stochastic motion in the sense that bundles of neighboring
trajectories form localized structures similar to fluid vortices.

\bigskip

DTM reduces the problem of determining the statistical behavior of the
stochastic trajectories to the calculation of weighted averages of smooth,
deterministic trajectories obtained from the Eulerian correlation of the
potential \cite{Vlad1998}. DTM is a semi-analytical statistical approach
that satisfies the statistical consequences of the invariance of the
potential.

The main idea of our approach is to study the stochastic equations (\ref{1})
in subensembles of realizations of the stochastic field, which contain all
realizations that have the same values of the stochastic potential and
velocity in the starting point of the trajectories 
\begin{equation}
(S):\quad \phi (\mathbf{0},0)=\phi ^{0},\quad \mathbf{v}(\mathbf{0},0)=%
\mathbf{v}^{0}.  \label{2}
\end{equation}%
A average potential $\Phi ^{S}$ and velocity $\mathbf{V}^{S}$ exist in each
subensemble. They are space-time dependent functions determined by the EC of
the potential as%
\begin{equation}
\Phi ^{S}(\mathbf{x},t)=\phi ^{0}\frac{E(\mathbf{x},t)}{E(0,0)}+v_{x}^{0}%
\frac{E_{y}(\mathbf{x},t)}{V_{x}^{2}}-v_{y}^{0}\frac{E_{x}(\mathbf{x},t)}{%
V_{y}^{2}},  \label{Fis}
\end{equation}%
\begin{equation}
\mathbf{V}^{S}(\mathbf{x},t)=\mathbf{e}_{z}\times \mathbf{\nabla }\Phi ^{S}(%
\mathbf{x},t)  \label{VS}
\end{equation}%
where $\mathbf{x}=(x,~y),$ $\phi ^{0},$ $v_{i}^{0}$ are the parameters of
the subensemble and the subscripts represent derivatives ($E_{i}\equiv
\partial E/\partial x_{i}).$ The amplitudes of the velocities are $%
V_{x}^{2}=-E_{yy}(\mathbf{0}),$ $V_{y}^{2}=-E_{xx}(\mathbf{0}).$

The stochastic equations are studied in each subensemble (S), where
trajectories with a high degree of similarity are obtained due to the
supplementary initial conditions (\ref{2}). Neglecting the fluctuations of
the trajectories, the average trajectory in (S) (\emph{the decorrelation
trajectory}) is obtained by averaging Eqs. (\ref{1}) in (S). One obtains
average equations with the same structure as the equations in each
realization (\ref{1}), but having the stochastic potential replaced by the
subensemble (conditional) average potential%
\begin{equation}
\frac{d\mathbf{X}^{S}}{dt}=\mathbf{e}_{z}\times \mathbf{\nabla }\Phi
^{S}+V_{d}\mathbf{e}_{y}.  \label{dectr}
\end{equation}%
This approximation is validated in \cite{Vlad2004}, where it is shown that
the DTM is the first order in a systematic expansion, and that the results
obtained in the second order are close to those of the first order.

The Lagrangian correlations are obtained as averages over the decorrelation
trajectories, by summing the contribution of each subensemble (S), weighted
by the probability that a realization belongs to the subensemble. In
particular, the LVC is approximated using the decorrelation trajectories as%
\begin{equation}
L_{ii}(t)\cong \int d\phi ^{0}P(\phi ^{0})\int
dv_{1}^{0}dv_{2}^{0}P(v_{1}^{0})P(v_{1}^{0})\ v_{i}^{0}\frac{dX_{i}^{S}(t)}{%
dt},  \label{LVC}
\end{equation}%
where $P$\ is the Gaussian probability 
\begin{equation*}
P(\phi ^{0})=\frac{1}{\sqrt{2\pi }\beta }\exp \left( -\frac{(\phi ^{0})^{2}}{%
2\beta ^{2}}\right) ,\ \ P(v_{i}^{0})=\frac{1}{\sqrt{2\pi }V_{i}}\exp \left(
-\frac{(v_{i}^{0})^{2}}{2V_{i}^{2}}\right) .
\end{equation*}%
\ \ 

Thus, the DTM is essentially an analytical method, which determines the
diffusion coefficients from a set of simple, deterministic trajectories that
result from the EC of the stochastic field. However, in most cases, the DT's
have to be numerically integrated. A computer code was developed for
calculating the time dependent diffusion coefficients. It determines the
DT's (\ref{dectr}) for a large enough number of subensembles and performs
the integrals in Eq.(\ref{LVC}). These computer calculations are at PC
level, for times of the order of 10 minutes.

\section{\protect\bigskip Effects of trajectory trapping on test particle
statistics}

\subsection{Memory effects and anomalous diffusion coefficients}

The Lagrangian velocity correlation (LVC) determines the time dependent
diffusion coefficient (\ref{D}) and the mean square displacement (\ref{MSD}%
). It is also a measure of the statistical memory of the stochastic motion.

The presence of trajectory trapping leads to a specific shape of the LVC for
a frozen turbulence $\phi (\mathbf{x})$. This function decays to zero in a
time of the order $\tau _{fl}$ but at later times it becomes negative, it
reaches a minimum and then it decays to zero having a long, negative tail.
The tail has a power law decay with an exponent that depends on the EC of
the potential \cite{Vlad2004b}. The positive and negative parts compensate
such that the integral of $L(t),$ the running diffusion coefficient $D(t)$,
decays to zero. The transport in such two-dimensional potential is thus
subdiffusive. The tail of the LVC shows that the stochastic trajectories in
frozen turbulence have long time memory. Thus, the LVC for incompressible
two-dimensional velocity fields is completely different from the LVC of
diffusion without trapping. The latter is decaying from $V^{2}$ to zero in a
time of the order $\tau _{fl}.$

This stochastic process is unstable in the sense that any weak perturbation
produces a strong influence on the transport. A perturbation represents a
decorrelation mechanism and its strength is characterized by a decorrelation
time $\tau _{d}.$ The weak perturbations produce long decorrelation times, $%
\tau _{d}\gg \tau _{fl}.$ In the absence of trapping, such a weak
perturbation does not produce a modification of the diffusion coefficient
because the LVC is zero at $t>\tau _{fl.}$\ In the presence of trapping,
which is characterized by long time LVC, such perturbation influences the
tail of the LVC and destroys the equilibrium between the positive and the
negative parts. Consequently, the diffusion coefficient is a decreasing
function of $\tau _{d}.$ It means that when the decorrelation mechanism
becomes weaker ($\tau _{d}$ increases) the transport decreases. This is a
consequence of the fact that the long time LVC is negative. This behavior is
completely different from that obtained in stochastic fields that do not
produce trapping. In this case, the transport is stable to the weak
perturbations, because the LVC is zero for $t>\tau _{fl}.$ An influence of
the decorrelation can appear only when the perturbation is strong such that $%
\tau _{d}<\tau _{fl}$ and it determines the increase of the diffusion
coefficient with the increase of $\tau _{d}.$

The decorrelation can be produced for instance by the time variation of the
stochastic potential, which destroys the Lagrangian correlations at $t>\tau
_{c}.$ The transport becomes diffusive with an asymptotic diffusion
coefficient that has the trapping scaling (\ref{dtr}), and thus it is a
decreasing function of $\tau _{c}$. This behavior is determined by the fact
that a stronger perturbation (with smaller $\tau _{d})$ liberates a larger
number of trapped trajectories, which contribute to the diffusion. For other
types of perturbations, the interaction with the trapping process produces
more complicated nonlinear effects. For instance, particle collisions lead
to the generation of a positive bump on the tail of the LVC \cite{V00} due
to the property of the 2-dimensional Brownian motion of returning in the
already visited places.

The average velocity $V_{d}\mathbf{e}_{y}$ does not provide a decorrelation
mechanism, but it determines an average potential that changes the structure
of the contour lines by producing bunches of opened lines (as discussed in
Section 2.1). A strong modification of the diffusion coefficient is produced
by the average velocity in the presence of trajectory trapping ($K_{\ast }>1$
and $K>1),$ which consists of a large increase of the parallel diffusion
\thinspace $\emph{D}_{y}$ and of the decrease of the perpendicular diffusion
\thinspace $\emph{D}_{x}$ \cite{Vlad2001}. Also, an interesting process of
"acceleration" appears. Since a fraction of the trajectories $n_{tr}$ are
trapped, they cannot participate to the average flux, which is produced only
by the free trajectories. The velocity field has zero divergence, thus the
distribution of the Lagrangian velocity is time independent, and in
particular, the average Lagrangian velocity is equal to $V_{d}.$ Due to
this, the average velocity of the free particles $V_{f}$ has to be lager
that $V_{d}$\ such that $n_{f}V_{f}=V_{d}.$\ Here $n_{f}=1-n_{tr}$ is the
fraction of free trajectories. The increased velocity $V_{f}$ is determined
by the selection of the stochastic velocity along the opened contour lines
of the potential, which are situated preferentially in regions where the
Eulerian velocity is oriented along $V_{d}\mathbf{e}_{y}.$ The trapping
determines in these conditions a strong modification of the distribution of
displacements, which elongates in the direction of $V_{d}\mathbf{e}_{y}$ and
eventually it splits in two parts (see Figure 2).

\begin{figure}[tbp]
\centerline{\includegraphics[height=.30\textheight]{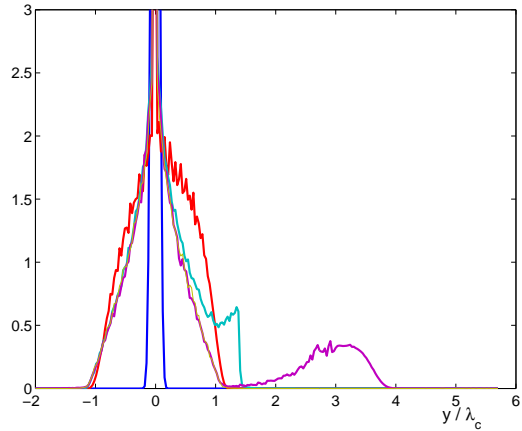}}
\caption{The probability of displacements along the average velocity $V_{d}%
\mathbf{e}_{y}$ obtained with the DTM for $K=10$ and $r=V_d/V=0.2$ at
several time moments.}
\end{figure}

The case of a stochastic potential that moves with the velocity $V_{d}%
\mathbf{e}_{y}$ is obtained by a Galilean transformation. This leads to
opposite flows: trapped particles move with the potential and the free ones
move in the opposite direction such that the total flux is zero, $%
n_{tr}V_{d}+n_{f}V_{f}=0.$

\subsection{Coherence and trajectory structures}

The statistical characteristics of the trapped and of the free trajectories
were studied in \cite{Vlad2004}. We have shown that the two types of
trajectories have completely different statistical characteristics.

The trapped trajectories have a quasi-coherent behavior. Their average
displacement, dispersion and probability of displacements saturate in a time 
$\tau _{s}$. The time evolution of the square distance between two
trajectories $\left\langle \delta x^{2}(t)\right\rangle $ is very slow
showing that neighboring particles have a coherent motion for a long time,
much longer than $\tau _{s}$. They are characterized by a strong clump
effect with the increase of $\left\langle \delta x^{2}(t)\right\rangle $
that is slower than the Richardson law. These trajectories form
quasi-coherent structures which are similar to fluid vortices and represent
eddying regions.

In time dependent potentials (with finite $\tau _{c}),$ the structures with $%
\tau _{s}>\tau _{c}$ are destroyed and the corresponding trajectories
contribute to the diffusion process. These free trajectories have a
continuously growing average displacement and dispersion. They have
incoherent behavior and the clump effect is absent. The probability of short
time displacements are non-Gaussian for both types of trajectories.

\begin{figure}[tbp]
\centerline{\includegraphics[height=.30\textheight]{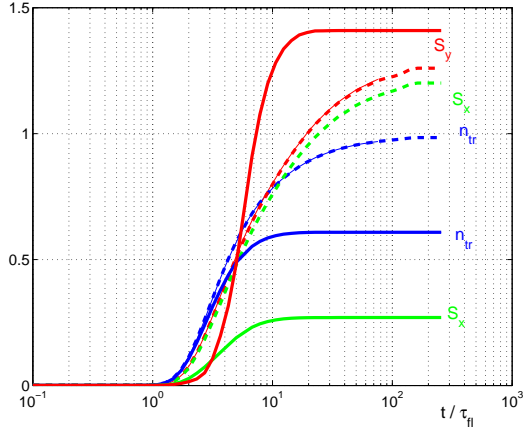}}
\caption{The formation of quasi-coherent trajectory structures in a time
independent stream function ($K,\protect\tau_c=\infty$). The time evolution
of $n_{tr}$, $S_{x}$, $S_{y}$ for $V_{d}=0$ (dashed lines) and $V_{d}=0.2$
(continuous lines). }
\end{figure}

\bigskip

The study of vorticity separation (presented in Section 4.1) requires more
information on the trajectory structures than obtained in \cite{Vlad2004}.
It is necessary to determine the fraction of trapped trajectories $n_{tr}$\
and the average size of the structures $S_{i}$\ as functions of the average
velocity $V_{d}.$\ 

These statistical quantities are obtained using the DTM as weighted averages
of the decorrelation trajectories $\mathbf{X}^{S}(t)$ in the static
potential. The solutions of Eq. (\ref{dectr}) are periodic functions in this
case with the periods $T(\phi ^{0},\mathbf{v}^{0})$ that depend on the
subensemble. The fraction of trapped trajectories at time $t$ is%
\begin{equation}
n_{tr}(t)=\int d\phi ^{0}P(\phi ^{0})\int
dv_{1}^{0}dv_{2}^{0}P(v_{1}^{0})P(v_{1}^{0})\ c_{tr}(t;\phi ^{0},\mathbf{v}%
^{0}),  \label{ntr}
\end{equation}%
where $c_{tr}(t;\phi ^{0},\mathbf{v}^{0})=1$ if $t>T(\phi ^{0},\mathbf{v}%
^{0})$\ and $c_{tr}(t;\phi ^{0},\mathbf{v}^{0})=0$ if $t<T(\phi ^{0},\mathbf{%
v}^{0}).$\ The sizes of the trajectory structures on the $i=x,y$\ directions
are%
\begin{equation}
S_{i}(t)=\int d\phi ^{0}P(\phi ^{0})\int
dv_{1}^{0}dv_{2}^{0}P(v_{1}^{0})P(v_{1}^{0})\ X_{i}^{\max }(t;\phi ^{0},%
\mathbf{v}^{0}),  \label{Si}
\end{equation}%
where $X_{i}^{\max }(t;\phi ^{0},\mathbf{v}^{0})=\max (X_{i}^{S})$ if $%
t>T(\phi ^{0},\mathbf{v}^{0})$ and $X_{i}^{\max }(t;\phi ^{0},\mathbf{v}%
^{0})=0$ if $t<T(\phi ^{0},\mathbf{v}^{0}).$ These functions describe the
growth of the trajectory structures.

The time evolution of the fraction of trapped trajectories $n_{tr}(t)$ and
the average size of the structures $S_{x}(t),$ $S_{y}(t)$ are plotted in
Figure 3 for $V_{d}=0$ (dashed lines) and for $V_{d}=0.2~$(continuous
lines). For $V_{d}=0,$ the structures continuously grow due to the large
size stream lines that lead to large periods of the corresponding
trajectories. An average velocity generates bunches of opened lines of the
total potential and limits the size of the islands of closed contour lines.
The size of the structures and $n_{tr}(t)$ saturate at finite $V_{d},$ as
seen in Figure 3. The average velocity determines the decrease of $n_{tr}$
and of the size of the structures in the perpendicular direction $S_{x},$
while the parallel size $S_{y}$ is increased. The structures are elongated
by the velocity.

\begin{figure}[tbp]
\centerline{\includegraphics[height=.30\textheight]{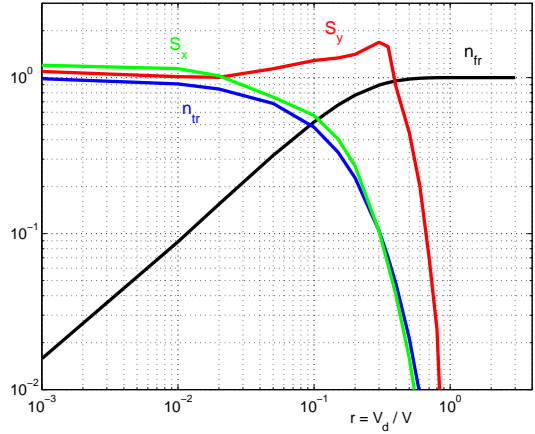}}
\caption{The dependence of the parameters of the quasi-coherent structures
on the average velocity. }
\end{figure}

The saturation values of the parameters of the structures $n_{tr},$ $S_{x},$ 
$S_{y}$\ and $n_{f}=1-n_{tr}$ are plotted in Figure 4 as functions of $%
V_{d}. $ One can see that the trajectory structures exists only for small
average velocities with $r=V_{d}/V<1$ (or $K_{\ast }>1).$ When $r\gtrsim 1,$
the structures are destroyed by the average velocity, which opens all the
contour lines of the total potential.\ 

\section{Effects of trapping on turbulence evolution}

The idea of the strong relation between tracer trapping or eddying and
turbulence evolution is sustained by analyzing two completely different
physical systems: the relaxation of turbulence in two-dimensional ideal
fluids and the drift turbulence in magnetically confined plasmas. Fluids are
stable and turbulence is externally produced by steering, while
inhomogeneous plasmas are unstable and generate spontaneously turbulence.
However, both systems are characterized by the inverse cascade of the energy
in the evolution of turbulence and by the tendency to self-organization. We
show that in both cases these processes are determined by trajectory
trapping, but with different mechanisms, which are essentially determined by
the different nonlinearities of the systems. The case of fluid turbulence is
analyzed in Section 4.1, where the statistics of the trajectories of the
vorticity elements in the neighborhood of a large vortex is studied. Section
4.2 contains a short review of recent results on the evolution of the drift
turbulence in magnetically confined plasmas. It is introduced in order to
underline the idea that trajectory trapping has the main role in both cases,
despite the fact that the physical processes that determine the inverse
cascade are completely different.

\subsection{Ideal fluids}

Ideal fluids are described by the Euler equation%
\begin{equation}
\partial _{t}\omega +\mathbf{v}\cdot \mathbf{\nabla }\omega =0,
\label{Euler}
\end{equation}

\begin{equation}
\mathbf{v}=\mathbf{e}_{z}\times \mathbf{\nabla }\phi ,  \label{vdr}
\end{equation}%
\begin{equation}
\omega =\bigtriangleup \phi ,  \label{voti}
\end{equation}%
where $\omega $ is the vorticity, $\phi $ is the stream function and $%
\mathbf{v}$ is the fluid velocity. Thus vorticity elements are advected by
the velocity field and the vorticity is conserved on the trajectories. The
nonlinearity of the process is determined by the relations (\ref{vdr})-(\ref%
{voti}) between $\mathbf{v}$ and $\omega ,$ which show that vorticity is an
active field.\ 

Numerical simulations have shown that the vorticity field evolves to
organized states characterized by isolated vorticity peaks with large
amplitudes. A process of separation of positive and negative vorticity and
vorticity concentration occurs during the relaxation of turbulent states.
Our aim is to understand how is possible to appear vorticity separation in a
turbulent state.

In order to study this tendency, we consider a turbulent state as initial
condition of Eq. (\ref{Euler}) that is represented by a stochastic stream
function with Gaussian distribution and with the Eulerian correlation $E(%
\mathbf{x}).$ We have taken for the results presented in the figures%
\begin{equation}
E(\mathbf{x})=E(r)=\frac{2-5r^{2}+r^{4}}{2}\exp \left( -\frac{r^{2}}{2}%
\right) ,  \label{Ecalc}
\end{equation}%
where $r=\left\vert \mathbf{x}\right\vert /\lambda .$ This function is
normalized to $E(\mathbf{0})=1$\ and its space integral is zero. The latter
condition is due to the choice of the total vorticity equal to zero. The
presence of a large scale vortex is modeled by an average stream function $%
\left\langle \phi (\mathbf{x})\right\rangle =xV_{d}$\ that accounts for an
average velocity $V_{d}\mathbf{e}_{y}.$ We consider first an uniform
velocity, and the effect of a small transverse gradient of $V_{d}\mathbf{e}%
_{y}$ is discussed at the end of this section.

Trajectory trapping occurs when $K_{\ast }>1$ and $K>1,$ as discussed in
Section 3.1. The first condition implies large amplitudes of the stochastic
component of the velocity compared to the average velocity $V>V_{d}.$\ The
second condition corresponds to large correlation times of the stochastic
stream function $\tau _{c}>\tau _{fl}=\lambda _{c}/V.$ The temporal
evolution of the stream function is determined in ideal fluids by the motion
of the vorticity elements, which modifies the vorticity field. This process
is slow compared to the time of flight, which means that fluid turbulence
has $K>1.$ Thus, trapping occurs in two-dimensional fluid turbulence when
the average velocity is zero or small ($V_{d}<V).$

The stream function and the vorticity are correlated and their two-point
correlation function results from $E(\mathbf{x})$ using Eq. (\ref{voti})\ 
\begin{equation}
E_{\phi \omega }(\mathbf{x})\equiv \left\langle \phi (\mathbf{x}^{\prime })\
\omega (\mathbf{x}^{\prime }+\mathbf{x})\right\rangle =\bigtriangleup E(%
\mathbf{x})  \label{cfiom}
\end{equation}%
Thus, the stream function and the vorticity in the same point $(\mathbf{x}=%
\mathbf{0})$ have opposite signs since $E(\mathbf{0})$ is the maximum of $E(%
\mathbf{x})$ and $\bigtriangleup E(\mathbf{0})<0.$ This correlation,
determined by the nonlinearity of the fluid flow, has an important role in
vorticity separation.

The trajectories of the vorticity elements are solutions of Eq. (\ref{1}).
Since they are independent on the vorticity and on its sign, it is not
evident how vorticity separation can occur.

We show that the average velocity $V_{d}\mathbf{e}_{y}$ determines two
transverse flows with opposite directions. They appear in the presence of
eddying and their direction is correlated with the sign of the stream
function corresponding to the stream line on which the trajectory is
trapped.\ \ 

The average displacement of the vorticity elements is determined using DTM
by summing the contributions of all subensembles. An equation similar to (%
\ref{LVC}) is obtained

\begin{equation}
\left\langle \mathbf{x}(t)\right\rangle =\int d\phi ^{0}~P(\phi ^{0})\int d%
\mathbf{v}^{0}~P(\mathbf{v}^{0})~\mathbf{X}^{S}(t),  \label{xmed}
\end{equation}%
\ where $\mathbf{X}^{S}(t)$\ is the decorrelation trajectory in the
subensemble $S$ determined from Eq. (\ref{dectr}). The component
perpendicular on the average velocity is zero. Actually it is of the order $%
10^{-5},$\ which is the accuracy of the numerical calculation. This is the
correct result for an incompressible velocity field. The average Lagrangian
velocity has to be equal to the Eulerian average velocity, which has no
component in the $\mathbf{e}_{x}$ direction. However, the conditional
average displacement for a given value of the initial stream function $\phi
^{0}$\ 
\begin{equation}
\left\langle x(t)\right\rangle _{\phi ^{0}}=P(\phi ^{0})\int d\mathbf{v}%
^{0}~P(\mathbf{v}^{0})~X^{S}(t)  \label{xmedfi}
\end{equation}%
is not zero, but it is an anti-symmetrical function of $\phi ^{0},$\ as
shown in Figure 5. One can see that the displacements are correlated with
the stream function and that $\left\langle x(t)\right\rangle _{\phi ^{0}}$
and of $\phi ^{0}$\ have the same sign. The displacements are negligible at
small time when there is no trapping. As time increases, the maximum of $%
\left\vert \left\langle x(t)\right\rangle _{\phi ^{0}}\right\vert $\
increases and eventually saturates.

Due to the anti-correlation of the stream function and vorticity in the same
point ($E_{\phi \omega }(\mathbf{0})<0),$ the positive peaks of the stream
function correspond to negative vorticity, while the negative ones
correspond to positive vorticity. Thus the correlation of the perpendicular
displacement with the stream function corresponds to a correlation between
displacements and vorticity.

\begin{figure}[tbp]
\centerline{\includegraphics[height=.30\textheight]{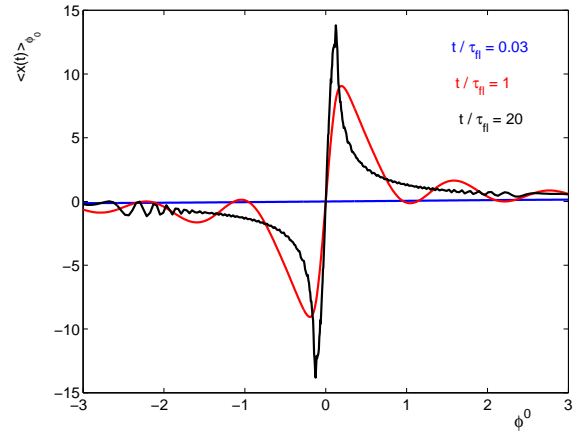}}
\caption{The average perpendicular displacement as function on the initial
stream function for $r=V_d/V=0.2$ and $K=\infty$. }
\end{figure}

\begin{figure}[tbp]
\centerline{\includegraphics[height=.30\textheight]{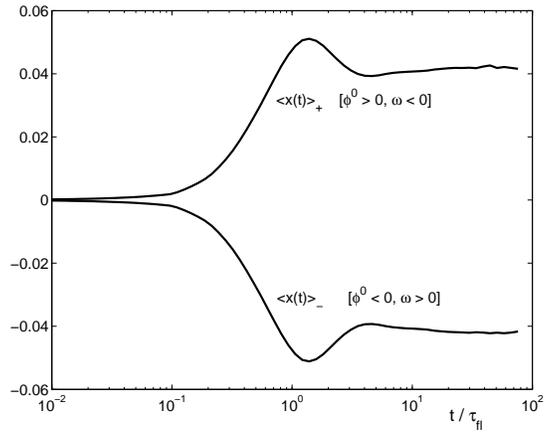}}
\caption{The average perpendicular displacements for the positive and
negative vorticity ($r=V_d/V=0.2$ and $K=\infty$). }
\end{figure}

The total average displacements for positive $\phi ^{0},$ $\left\langle
x(t)\right\rangle _{+},$ and respectively for negative $\phi ^{0},$ $%
\left\langle x(t)\right\rangle _{-},$ are represented in Figure 6. They are
symmetrical and compensate, as required by the zero divergence of the
velocity field. These average displacements determine opposite average
velocities in the case of time dependent vorticity fields 
\begin{equation}
V_{s+}=\frac{\left\langle x(\tau _{c})\right\rangle _{+}}{\tau _{c}},\ \
V_{s-}=\frac{\left\langle x(\tau _{c})\right\rangle _{-}}{\tau _{c}}=-V_{s+},
\label{vspm}
\end{equation}%
where $\tau _{c}$ is the correlation time.\ 

In terms of vorticity, the negative vorticity elements that are trapped on
the stream lines with $\phi ^{0}>0$ move with the velocity $V_{s+}$ in the
direction of the gradient of the large scale stream function. The positive
vorticity elements have an opposite average velocity. Thus, the positive and
negative vorticity elements that evolve on trapped trajectories separate
with a speed $V_{s}=V_{s+}-V_{s-}=2V_{s+}.$ This speed is plotted in Figure
7 as function of the Kubo number $K=\tau _{c}/\tau _{fl}.$ One can see that $%
V_{s}$ is maximum for $K$\ of the order one, and it decays at $K\gg 1$\ as $%
K^{-1}$ due to the saturation of the average displacements. The maximum
value is smaller than $V_{d},$\ but not negligible.\ 

\begin{figure}[tbp]
\centerline{\includegraphics[height=.30\textheight]{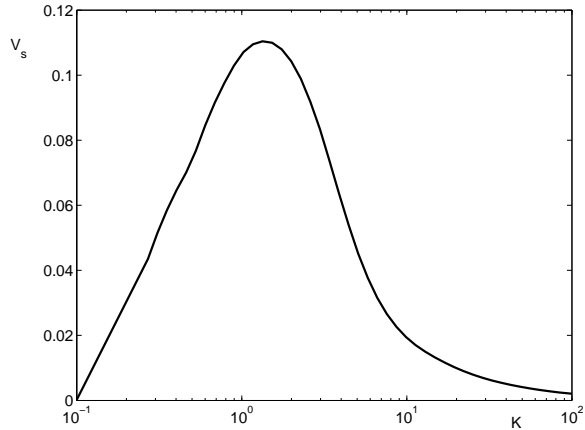}}
\caption{The average normalized velocity of positive and negative vorticity
separation, $V_s/V$ as function of the Kubo number for $r=0.2$. }
\end{figure}

The physical mechanism for the vorticity separation consists of the
influence of the average velocity on the small structures of trapped
trajectories. $V_{d}$ determines a difference between the average velocities
on the two sides of the closed trajectories that are oriented along $V_{d}%
\mathbf{e}_{y}.$ The velocity is $\partial _{x}\phi +V_{d}$ on one side and $%
\partial _{x}\phi -V_{d}$ on the other side. The difference of velocities
leads to the accumulation of the vorticity elements on the small velocity
side and to the increase of the vorticity in that region. As the direction
of the velocity on the contours of $\phi $ depend on the sign of $\phi ,$\
the accumulation of the vorticity elements occurs on opposite sides for
contours corresponding to $\phi $ and $-\phi .$\ The time variation of the
velocity field destroys the trajectory structures in a time of the order $%
\tau _{c},$ and the coherent displacements of the vorticity elements produce
transverse average velocities.

The average displacement $\left\langle y(t)\right\rangle $ along the average
velocity $V_{d}\mathbf{e}_{y}$ is produced, as explained in Section 3.1,
only by the free trajectories. They determine the elongation and eventually
the splitting of the distribution of displacements (see Figure 2). Thus, the
vorticity elements spread along $V_{d}\mathbf{e}_{y}$\ while they have
transverse drifts.\ \ 

The average flows of the vorticity elements determine the modification of
the vorticity distribution. Due to the correlation (\ref{cfiom}), this
changes the distribution of the stream function: the contour lines of the
stream function follow the average motion of the vorticity.

These findings are in agreement with the image provided by the numerical
simulations (see for instance \cite{McWilliams90}), which show the emergence
of large vortices that elongate and attract small vortices of the same sign.
We note that the process analyzed here is different from the vortex drift
transverse to a sheared flow analyzed in \cite{Schecter-Dubin} since it
involves vorticity elements in turbulent flows and not localized vortices.
In our case the drift is produced by an average velocity and the existence
of a gradient of the background vorticity is not necessary.

A vortex determines a velocity that has a gradient in the transverse
direction. The gradient length is much larger than the correlation length of
the stream function in the case of a large vortex in a small scale
stochastic velocity field. In these conditions, the effects of the gradient
can be estimated using the dependence on $V_{d}$\ of the statistical
characteristics obtained at constant $V_{d}.$\ 

The transverse drift of the small structures with the same sign of the
vorticity as the large vortex is in the direction of the gradient and brings
the structure toward larger $V_{d}.$ This determines the decrease of the
transverse dimension of the structure and the increase of its elongation, as
seen in Figure 4. When $V_{d}\gtrsim V,$ the small structures are destroyed.
The fraction of trapped trajectories and the size of the structures decrease
rapidly for $V_{d}>V.$ Their vorticity is included in the large vortex, and
it leads to the increase of the large scale energy (inverse cascade). On the
contrary, the structures with opposite sign of vorticity drift toward
smaller $V_{d},$ against the gradient. Their size increases and their
elongation is reduced.

In conclusion, a large size vortex in a decaying two-dimensional turbulence
attracts the small size vortices with the same sign of vorticity while the
small vortices of the opposite sign are repelled.\ This effect of vorticity
separation according to its sign is produced by trapping of the vorticity
elements combined with the existence of the correlation of the stream
function with the vorticity imposed by the nonlinearity (\ref{voti}).

\subsection{Drift turbulence}

The nonlinear stage in the evolution of drift type turbulence is essentially
a consequence of ion trajectory trapping or eddying in the structure of the
stochastic potential. This conclusion is drawn from a study of test modes on
turbulent plasmas \cite{Vlad2013}, \cite{VladSpineanu2013} based on a new
Lagrangian approach that extends the type of methods initiated by Dupree 
\cite{Dupree} to the nonlinear regime characterized by trapping. Drift waves
are unstable due to the electron kinetic effects that produce the
dissipation mechanism to release the energy, combined with the ion
polarization drift \cite{Goldstone}. Beside this, the polarization drift has
a more complex influence determined by its nonzero divergence, which
produces a weak compressibility effect in the background turbulence.

Test modes on turbulent plasmas were studied for drift turbulence in
constant magnetic field starting from the basic description provided by the
drift kinetic equations. Analytical expressions are derived, which
approximate the growth rate $\gamma (\mathbf{k})$ and the frequency $\omega (%
\mathbf{k})$ of a test mode with wave number $\mathbf{k}=(k_{x},~k_{y})$ as
functions of the characteristics of the background turbulence. These
functions provide the tendency of turbulence evolution.

The dispersion relation of the test modes in turbulent plasma is shown to be
the same as in quiescent plasma, except for a time dependent function $M(t)$%
, which embeds all the effects of the background turbulence. This function
appears in the propagator of the test mode, which is obtained using the
formal solution of the ion equation with the method of characteristics. The
characteristics are ion trajectories in the stochastic background potential.
They are described by equations similar to Eq. (\ref{1}), with the
difference that the potential $\phi $ is moving with the effective
diamagnetic velocity $V_{d}\mathbf{e}_{y}$, which leads to the replacement
of $y$\ with $y^{\prime }=y-V_{d}~t.$\ The function $M(t)$ is evaluated
using the DTM, and the dispersion relation is analytically solved.\ \ 

Drift turbulence develops in the initial stage on a wide range of wave
numbers. Ion trajectories are not trapped at these small amplitudes of the
background turbulence and they have Gaussian distribution. Their diffusion
determines the damping of the large wave number modes. The correlation
lengths $\lambda _{i}$ remain close to the ion Larmor radius $\rho _{i}$
during this stage. Turbulence amplitude $\beta $ increases and the shape of
the EC is not changed.

When the amplitude reaches values that make $K_{\ast }>1,$ ion trajectory
trapping appears and generates vortical structures of trapped ions as
discussed in Section 3.2. They determine the decrease of the frequencies,
which leads to the decrease of $\gamma $ and to the displacement of the
unstable range of wave numbers to smaller values. The maximum of the growing
rate is displaced from $k_{y}\cong 1/\rho _{i}$ to $k_{y}\cong 1/S,$ where $%
S $ is the average size of trajectory structures. Turbulence amplitude
continues to increase in this stage, but with a smaller rate. This
determines the increase of the fraction of the trapped ions $n_{tr}$ and of
the size $S$ of the trajectory structures.

Large scale correlations are generated through a nonlinear process of
decreasing the wave number of the unstable modes. In the same time, the
diffusive damping produced by the free trajectories continues to act on the
large $\mathbf{k}$ modes, which are attenuated. This process of inverse
cascade of turbulence energy stops well before the size of the potential
cell becomes comparable with the system size. A different effect appears at
larger trapping rates, which is also connected with trapping.

When the fraction of trapped ions $n_{tr}$ becomes comparable with the
fraction of free ions $n_{f},$ ion flows are generated by the trapping
process as explained in Section 3.1. Trapped ions move with the potential,
while free ions move in the opposite direction such that the total flux is
zero. This determines the splitting of the distribution of ion displacement
(as in Figure 2). As shown in \cite{Vlad2013}, an essential change of test
modes is produced. The drift modes are damped and a different type of modes,
zonal flow modes, are generated in this strongly nonlinear regime. They have 
$k_{y}=0$ and very small frequencies and are produced by the combined action
of the ion flows and of the compressibility due to the polarization drift in
the background turbulence. The damping of the drift modes is not determined
by the zonal flow modes as generally believed. There is only an indirect
contribution through the diffusive damping, which is increased by the zonal
flow modes \cite{VladSpineanu2013}.\ The decay of the drift turbulence
determines the decrease of $n_{tr}$ and of the growth rate of the zonal flow
modes, which is proportional to $n_{tr}.$ Consequently, drift turbulence
does not saturate, but it oscillates between weak and strong trapping. The
predator-prey paradigm is not sustained by these results, although there is
time correlation between the growth of zonal flow modes and the damping of
the drift modes.

In conclusion, trapping determines in the first phase the increase of the
correlation length by displacing the range of unstable modes to small $k\sim
1/S.$ In the second stage, when it becomes stronger and produces ion flows
in the moving background turbulence, trapping leads to the nonlinear damping
of the drift modes. In the same time, trapping combined with the
polarization drift generates zonal flow modes.

\section{Conclusion}

Particle motion in two-dimensional incompressible stochastic velocity fields
is characterized by a mixture of random and quasi-coherent aspects,
represented by a random sequence of large jumps and trapping or eddying
events. Trajectory trapping generates quasi-coherent structures of
trajectories.

The development of statistical methods that are adequate for the description
of such complex trajectories has contributed to the understanding of the
nonstandard characteristics of the transport in these systems. Various
aspects of the turbulent transport were analyzed in the test particle
approach using these semi analytical methods. \ These studies are now in a
new stage that consists of developing Lagrangian methods for the study of
the evolution of the turbulence in the nonlinear regime. First results are
reported in \cite{Vlad2013} and in the present paper. They lead to the
conclusion that trajectory trapping has the essential role in turbulence
evolution. The physical mechanism depends on the specific nonlinearity of
each system, but it is related to trajectory trapping. It appears that the
quasi-coherent component of the trajectories determine the nonlinear effects
in turbulence while the random component provides a damping or dissipation
process.

\emph{\ }In the case of fluids, the trapped vorticity elements have an
average motion, which leads to average velocities with the directions
dependent on the vorticity sign. Then, due to the correlation $E_{\phi
\omega },$\ the stream function is also transported by the perpendicular
average flows. The effect of this process is the separation of the vorticity
with different signs since negative vorticity moves in the sense of the
stream function gradient and positive vorticity in the opposite sense. A
small gradient of the average velocity that characterizes a large scale
vortex determines the decrease of the attracted small vortices and
eventually the inclusion of their vorticity in the large vortex. This
process determines the flow of the energy from small to large scales.

In the case of drift turbulence, the trapped ions determine the displacement
of the wave numbers of the most unstable modes towards small values, while
the free ions contribute to the diffusive damping of the large $k$ modes.
The correlation length of the stochastic potential increases until the
process of nonlinear damping becomes dominant. Ion trapping has an important
role in this process and also in the generation of the zonal flow modes.

\bigskip

\textbf{Acknowledgments} This work was supported by the Romanian Ministry of
National Education under the contract PN 09 39 01 01.










\end{document}